\documentclass[oldversion]{aa}
\usepackage{graphicx}
\usepackage{txfonts}
\usepackage{ulem}

\begin{document}
   \title{Restless quiescence: thermonuclear flashes between transient X-ray outbursts}

   \author{E. Kuulkers\inst{1}
	  \and
          J.J.M. in 't Zand\inst{2}
          \and
	  J.-P. Lasota\inst{3,4}
	  }

   \authorrunning{E.~Kuulkers et al.}
   \titlerunning{Restless quiescence}

   \offprints{E. Kuulkers}

   \institute{ESA, European Space Astronomy Centre (ESAC), P.O.~Box 78, 28691, Villanueva de la Ca\~nada (Madrid), Spain
              \email{Erik.Kuulkers@esa.int}
          \and
              SRON Netherlands Institute for Space Research, Sorbonnelaan 2, 3584 CA Utrecht, The Netherlands
          \and
	      Institut d'Astrophysique de Paris, UMR 7095 CNRS, UPMC Univ Paris 06, 98bis Bd Arago, 75014 Paris, France
          \and
              Astronomical Observatory, Jagiellonian University, ul.\ Orla 171, 30-244 Krak\'ow, Poland
             }

   \date{Received; accepted}

  \abstract{
   For thermonuclear flashes to occur on neutron-star surfaces, fuel must have been accreted from a donor star. 
   However, sometimes flashes are seen from transient binary systems
   when they are thought to be in their quiescent phase, during which 
   no accretion, or relatively little, is expected to occur. We investigate 
   the accretion luminosity during several such flashes, including 
   the first-ever and brightest detected flash from \object{Cen\,X-4} in 1969. 
   We infer from observations and theory that
   immediately prior to these flashes the accretion rate must have been
   between about 0.001 and 0.01 times the equivalent of the Eddington limit,
   which is roughly 2 orders of magnitude less than the peak accretion rates seen
   in these transients during an X-ray outburst and 3--4 orders of magnitude 
   more than the lowest measured values in quiescence. 
   Furthermore, three such flashes, including the one from \object{Cen\,X-4}, occurred within 
   2 to 7 days followed by an X-ray outburst. A long-term episode of 
   enhanced, but low-level,
   accretion is predicted near the end of the quiescent phase by the disk-instability 
   model, and may thus have provided the right conditions for these flashes to occur.
   We discuss the possibility of whether these flashes acted as triggers of the outbursts, signifying 
   a dramatic increase in the accretion rate. Although it is difficult to rule
   out, we find it unlikely that the irradiance by these flashes is
   sufficient to change the state of the accretion disk in such a
   dramatic way.
}


   \maketitle
%

\section{Introduction}

In low-mass X-ray binaries (LMXBs) a neutron or black hole
accretes matter via an accretion disk from a less massive
Roche-lobe filling companion star. In many
LMXBs the accretion from the disk on the compact object is
transient. After a transient outburst (hereafter referred to as outburst), 
lasting typically a few months, the system settles in quiescence for a few
months to several decades. The disk-instability model (DIM;
Osaki 1974, see Lasota 2001 for a review)
predicts that, if in quiescence the disk extends down to the
stellar surface (or the last stable Keplerian orbit), the
accretion rate is negligibly low
($\sim$10$^{5}$\,g\,s$^{-1}$; Lasota et al.\ 2008, see
Eq.~(\ref{eq:mdotcrit}) in Sect.~4). However, in
general the X-ray emission of quiescent transient sources
corresponds to accretion rates that cannot be qualified as
negligible (e.g., van Paradijs et al.\ 1987, Campana et al.\ 2004). 

In the case of neutron-star transients there has been a
controversy on the origin of the quiescent X-ray luminosity. On
the one hand, quiescent-disk truncation easily explains the
luminosity (Lasota et al.\ 1996, Dubus et al.\ 2001, 
Narayan \&\ McClintock 2008), is also applicable to black-hole systems, and
has been confirmed by observations (e.g., Done 2002). On the
other hand, Brown et al.\ (1998) propose that quiescent X-rays
have their origin in heating of the neutron-star crust by
nuclear reactions and not in accretion. Both
models have their difficulties. The observed rapid variability
of the quiescent X-ray flux and the presence of a
substantial power-law component (as seen in, e.g., the LMXB
transient \object{Cen\,X-4}, see Campana et al.\ 1997, 2004, 
Rutledge et al.\ 2001), as well as a very low quiescent X-ray luminosity 
(less than about several times 10$^{30}$\,erg\,s$^{-1}$ for
\object{1H\,1905+000}; Jonker et al.\ 2007), are difficult to reconcile with the
deep crustal heating model (see, e.g., Jonker 2008 for a recent discussion).
The lack of a well understood
disk-truncation mechanism (see, however, Liu et al.\ 2002)
and the overpredicted ratio of neutron-star to
black-hole quiescent X-ray luminosities (Menou et al.\ 1999)
are the weaknesses of its competitor. Therefore, any independent
estimate of the quiescent accretion rate in transient systems
would be a great help in resolving the controversy.

Type I X-ray bursts (Grindlay et al.\ 1975, Belian et al.\ 1976, 
Hoffman et al.\ 1978) are thermonuclear flashes at the
surface of a neutron star (Joss 1977, Maraschi \&\ Cavaliere 1977, 
Lamb \&\ Lamb 1978; for reviews see, e.g., Lewin et al.\ 1993, 
Strohmayer \&\ Bildsten 2006). We hereafter refer to these events as flashes.
If the energy release
during a flash is fast and large enough, the local luminosity on the
neutron star surface can surpass the Eddington limit, resulting in a lift-up of the photosphere.
Such flashes are referred to as photospheric radius-expansion X-ray bursts.
During the expansion phase the inferred temperature decreases,
whereas the inferred emitted area increases (see, e.g., Lewin et al.\ 1993, for a review).

The first observed flash, in retrospect, was the event detected on 
July 7, 1969, with {\it Vela 5B} from \object{Cen\,X-4} (Belian et al.\ 1972; see Sect.~2). 
The event lasted for about 10\,min, and is still the
brightest ever observed with a peak flux of about
60\,Crab\footnote{The commonly used Crab unit is equivalent to
about 2$\times$10$^{-8}$\,erg\,cm$^{-2}$\,s$^{-1}$ in the
classical 2--10\,keV photon-energy band.} (3--12\,keV). Two
days after the flash \object{Cen\,X-4} went into an
X-ray outburst, which peaked at about 25\,Crab
(3--12\,keV) and lasted for about 80 days (Conner et al.\ 1969;
Evans et al.\ 1970; see also Sect.~2). Except for the flash,
no other X-ray emission was detected from \object{Cen\,X-4}
before the outburst (Belian et al.\ 1972).
Interestingly, a similar situation recently occurred in another
source: a $\simeq$40\,s long flash was detected
from \object{IGR\,J17473$-$2721} about 2 days prior to the 
X-ray outburst (Del Monte et al.\ 2008, Markwardt et al.\ 2008). 
In an other case three flashes
separated by 2--3~days were seen during the beginning of an
X-ray outburst of \object{2S\,1803$-$245} (Cornelisse et al.\ 2007). 
Our inspection of the {\it RXTE} All Sky Monitor (ASM)
light curve shows that the first flash, which
lasted for about 40--50\,s, occurred when there was no
detectable X-ray emission. The 
following two flashes occurred when the source
showed X-ray emission at a slightly elevated level. About a
week later the source developed into a bright X-ray
outburst.

A few other LMXB transients have shown flashes
after and/or in between X-ray outbursts when their
X-ray emission was below the detection thresholds 
and the systems were inferred to be in their quiescent phase:
\object{2S\,1711$-$339}, \object{SAX\,J1808.4$-$3658} and \object{GRS\,1747$-$312}
(Cornelisse et al.\ 2002a, in 't Zand et al. 2001, 2003b,
respectively). Also \object{Cen\,X-4} may have shown a
flash $\sim$2 years after its 1969 X-ray
outburst (Gorenstein et al.\ 1974; see Appendix \ref{apollo}).
Related examples are various flashes seen from
sources without detectable pre-flash emission, the so-called
burst-only sources (Cocchi et al.\ 2001, Cornelisse et al.
2002a,b, and references therein).

For all the afore-mentioned flashes, accretion must
have been ongoing prior to these events at a level that
is orders of magnitude lower than that achieved during
an outburst, but, interestingly, higher than quiescent levels.
This brings about prospects for new
constraints on models for the emission during the quiescent phase. We explore in Sect.~3
the observed and expected mass-accretion rates around the 
times of the flashes seen for the sources
presented above. Furthermore, the flashes which are
within 2--7~days followed by X-ray outbursts suggest
the presence of a physical process which has thus far not been
discussed in the literature\footnote{Belian et al.\ (1972)
called \object{Cen\,X-4}'s 1969 flash a probable precursor to its
subsequent X-ray outburst. They suggested the two
events to be associated, but no physical scenario was
discussed.}: that the flashes serve as triggers for
accretion-disk instabilities resulting in X-ray
outbursts. We study in Sect.~4 the viability of this idea in the context
of \object{Cen\,X-4}.

\section{\object{Cen\,X-4}: revisiting old data}

\begin{figure*}[top]
\centering
  \includegraphics[height=.3\textheight,angle=-90]{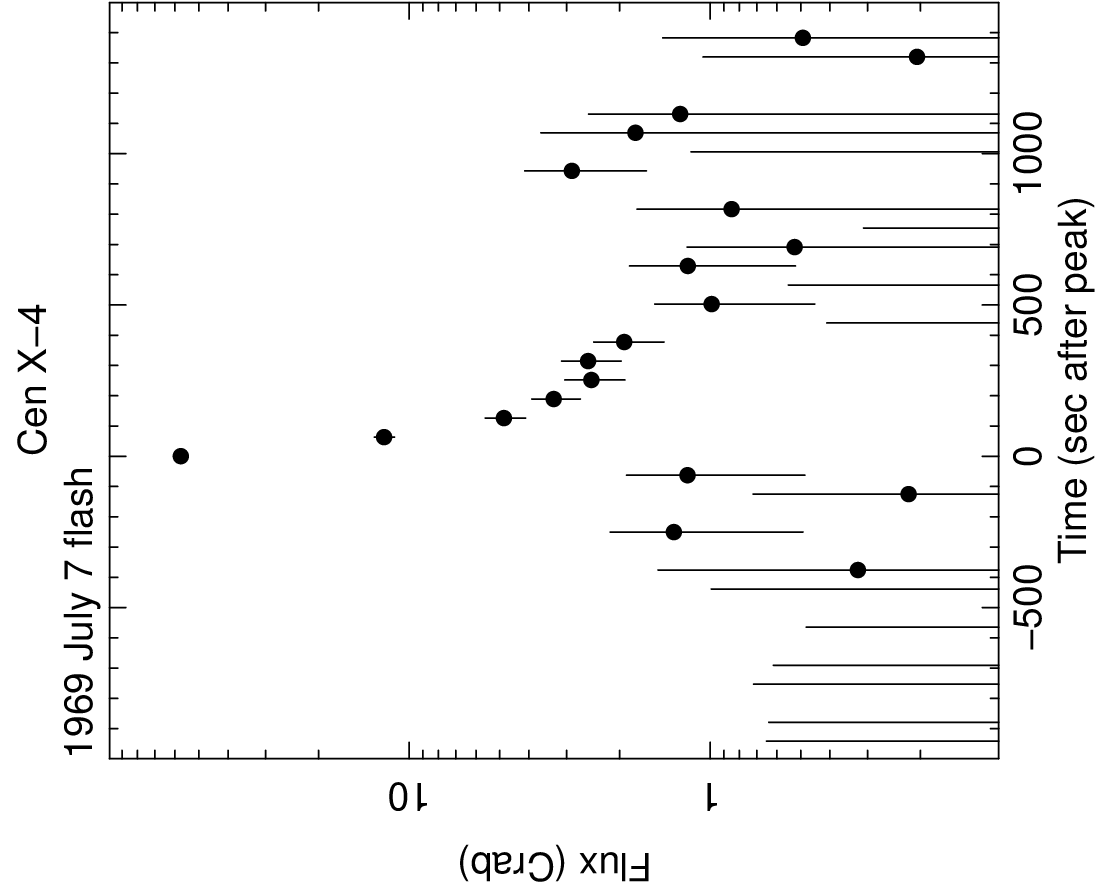}
\hspace{0.3cm}  \includegraphics[height=.4\textheight,angle=-90]{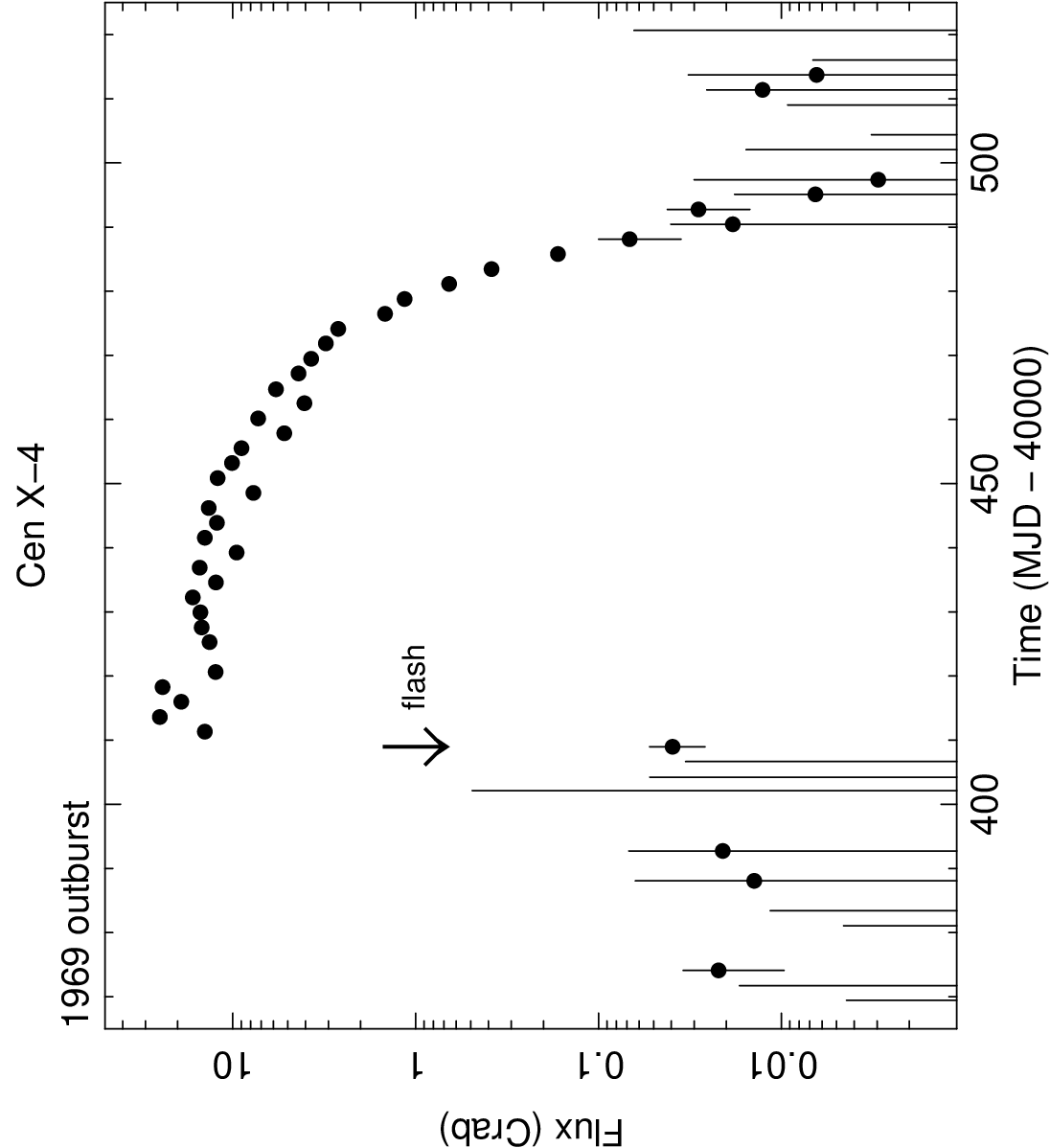}
  \caption{{\it Left:} {\it Vela 5B} light curve (3--12\,keV) of the X-ray event
from \object{Cen\,X-4} which occurred on July 7, 1969. Time = 0\,s
corresponds to UT 01:56:56. See also Belian et al.\ (1972).
{\it Right:} {\it Vela 5B} light curve (3--12\,keV) of the
X-ray outburst of \object{Cen\,X-4} which lasted from 1969,
July-September. See also Evans et al.\ (1970). Data are shown
from near the start (1969 May 28) up to about 155 days after
the start (1969 Oct 29) of the Vela~5B mission.
The time of the flash is indicated with an arrow.
}
\end{figure*}

\object{Cen\,X-4}'s 1969 flash was reported to reach a peak
flux\footnote{Whether the flux is bolometric or in the
3--12\,keV band is not entirely clear from Belian et al.\ (1972). 
Assuming the X-ray detectors onboard {\it Vela 5B} to be ideal detectors
(i.e., with a 100\%\ quantum efficiency), we can translate the quoted 3--12\,keV count rate
(2850\,c\,s$^{-1}$, Belian et al.\ 1972) into a photon flux, and
subsequently estimate the bolometric flux assuming a black body
with a temperature of 3$\times$10$^7$\,K (see Belian et al.\ 1972).
This results in a bolometric flux estimate of about 1.8$\times$10$^{-6}$\,erg\,cm$^{-2}$\,s$^{-1}$, i.e.,
higher than the quoted peak flux. Since the X-ray detectors are not ideal,
quantum efficiencies are lower, which results in even higher estimated bolometric flux values.
This and the fact that Belian et al.\ (1972) speak of an ``energy" flux,
suggests the quoted peak value to be the 3--12\,keV flux.
However, given the uncertainties involved and that the quoted peak flux is closer to
the bolometric peak flux of that observed during \object{Cen\,X-4}'s flash
seen with {\it Hakucho}, we use the quoted value as the bolometric flux.}
of about 1.4$\times$10$^{-6}$\,erg\,cm$^{-2}$\,s$^{-1}$.  The event rose
to an observed maximum within up to about 1\,min and lasted about
10\,min; the shape of the decay after the peak was more
consistent with a power-law than with an exponential (see also
below). During the decay the emission softened: at the peak it
could be described by a black-body with a temperature of about
2.6\,keV, whereas in the tail the temperature was about
1.3\,keV (Belian et al.\ 1972). We note that the effective
black-body radius at the peak would be about 6.5\,km (at
1.2\,kpc, see below). The energy release during the flash was
estimated to be about 5$\times$10$^{39}$\,erg (at 1.2\,kpc;
Matsuoka et al.\ 1980). It was recognized later that these are
the characteristics of a flash, although it was
noted that the event lasted unusually long (Fabbiano \&\
Branduardi 1979, Matsuoka et al.\ 1980, Kaluzienski et al.\
1980; see also Hanawa \&\ Fujimoto 1986). However, such long
flashes, although rare, have since been seen in
other LMXBs as well (see, e.g., Kuulkers 2004, in 't Zand et
al.\ 2007, Falanga et al.\ 2008).

The position of \object{Cen\,X-4} has been refined (Hjellming 1979,
Canizares et al.\ 1980) since the early {\it Vela 5B} reports;
we, therefore, decided to re-analyse the {\it Vela 5B} data.
The raw data files are archived at HEASARC (see Whitlock et
al.\ 1992, Whitlock \&\ Tyler 1994). {\it Vela 5B} (see Conner
et al.\ 1969) provided the first X-ray all-sky monitor; it was,
however, designed to be a nuclear-test detection satellite. The
satellite rotated about its nadir-fixed spin axis with a 64-s period, and
orbited the Earth in around 112\,hrs. The scintillation X-ray
detector (XC) was located at about 90$\degr$ from the spin
axis, and so covered the X-ray sky twice per satellite orbit.
Data were telemetered in 1-s count accumulations. The X-ray
detector provided data in two energy channels, 3--12 keV and
6--12 keV. A slat collimator limited the field of view to a
FWHM aperture of 6.1$\degr$$\times$6.1$\degr$; the
effective detector area was about 26\,cm$^2$.

We extracted the light curves, assuming that no other strong X-ray
sources were in the field of view. We used two time scales. For
the X-ray outburst we used 56\,hrs (i.e., half the {\it
Vela 5B} satellite orbital period), and included data up to the
recommended 5$\degr$ off-axis from the position of \object{Cen\,X-4}.
For the flash we used the 1-s info. The beginning
of the flash occurred at a larger off-axis angle than the
recommended value; here we therefore included data from up to 
6$\degr$. Data flagged for an unstable spin period and for
pointing errors were not included. The data were corrected
using a sinusoidally modelled background. For some sources,
removal of the fitted background leaves a slightly negative
average in the count rate; this is also the case for \object{Cen\,X-4}.
We corrected the 56-hrs average light curve for the observed
negative average in the count rate just before and after the
X-ray outburst (about $-$3\,c\,s$^{-1}$). The 1-s data are corrected
for the collimator response and the time stamps are corrected
to solar-system barycentric time. The errors on these data are
determined as follows: we assumed Poisson counting statistics
in the light curves uncorrected for background and collimator;
the errors were then propagated when all corrections were made.
The 56-hrs count rates are weighted averages of the 1-s data;
the errors take into account uncertainties introduced by
background removal and collimator response correction, as well
as counting statistics (see Whitlock et al.\ 1992). We normalized the count rates to the
{\it Vela 5B} Crab rate observed in the 3--12\,keV band between
1969, May 28 and Oct 29, i.e., around the time of the 1969 X-ray 
outburst (43.2$\pm$0.2\,c\,s$^{-1}$).

The flash (Fig.~1, left panel) rose to the
observed peak within up to 62\,s. We find a peak flux of
57$\pm$2\,Crab, which is more than about 3 orders of magnitude
above the upper limits on the pre-flash level (see Fig.~1,
right panel). Our revised peak intensity is a factor of
$\simeq$1.15 lower than that reported by Belian et al.\
(1972). Assuming the count rate and black-body flux scale
linearly, we infer a peak flux of about
1.2$\times$10$^{-6}$\,erg\,cm$^{-2}$\,s$^{-1}$. 
We only have sparse timing information. 
Because of the data gaps the actual peak
flux may have been even higher; we also
can not say with certainty whether the event showed a
photospheric radius-expansion phase indicative of 
(super-)Eddington fluxes from the flash
or not.

Until about 150\,s after the peak the exponential decay
time is about 44\,s. After that excess emission above the
expected exponential decay is observed. Such excess emission is
similar to that seen in other flashes from,
e.g., \object{Aql\,X-1} (Czerny et al.\ 1987), \object{X1905+000} (Chevalier \&\
Ilovaisky 1990) and \object{GX\,3+1} (Chenevez et al.\ 2006).
Due to this excess emission the shape of the full decay looks
more like a power law (see above).

The X-ray outburst (Fig.~1, right panel) reached a
peak of 25.0$\pm$0.3\,Crab, which is about 3 orders of
magnitude above the upper limits on the quiescent level.
Note that the light curve shows variations on the
56-hr time scale (see also Evans et al.\ 1970). This is
attributed to gain variations due to a 60$\degr$ satellite
temperature change from one side of the satellite orbit to the
other. This has not been (and can not be any more) taken into
account in the data reduction, because of the lack of
pre-launch tests, and the lack of information regarding the
temperature values in our time frame of interest (see Whitlock
et al.\ 1992).

A second X-ray outburst occurred in 1979, which
reached a peak flux of about 4\,Crab (3--6\,keV) and lasted
about a month (Kaluzienski et al.\ 1980). During the late
stages of the outburst a flash was observed with
{\it Hakucho} (Matsuoka et al.\ 1980). This flash reached a
peak flux of about 25\,Crab (1.5--12\,keV; with an estimated
uncertainty of about 5\%, M.~Matsuoka, private communication)
or a bolometric peak flux of 1$\times$10$^{-6}$\,erg\,cm$^{-2}$\,s$^{-1}$,
attained a peak black-body temperature of 2.5$\pm$0.8\,keV, lasted for $\simeq$100\,s, and
had an energy output of about 1$\times$10$^{39}$\,erg (at 1.2\,kpc;
Matsuoka et al.\ 1980). There is no clear evidence of a radius-expansion
phase in this flash either.

The peak fluxes of the 1969 and 1979 flashes have
been used to estimate the proximity of the system; the upper
limit was about 1.2\,kpc (Matsuoka et al.\ 1980, Kaluzienski et
al.\ 1980, Chevalier et al.\ 1989). Our revised 1969 
flash peak flux and the theoretical value for the
Eddington-limited luminosity, $L_{\rm Edd}$
($\simeq$2$\times$10$^{38}$\,erg\,s$^{-1}$)
\footnote{The Eddington luminosity as measured by a distant
observer is $L_{\rm Edd}$=$(4\pi cGM/\kappa )[1-2GM/(Rc^2)]^{1/2}$, where
$c$, $G$, $M$, $\kappa$ and $R$ are the speed of light, the gravitational constant,
the mass of the object, the electron scattering opacity and
the radius of the object, respectively (see, e.g., Lewin et al.\ 1993).}, 
for a 1.4\,M$_{\odot}$ star which accretes material with solar composition 
(see Sect.~3.1) translate again to an upper limit of about 1.2\,kpc (assuming
isotropic radiation). This upper limit is consistent with that
derived from other constraints, i.e., between 0.9\,kpc and
1.7\,kpc (Gonz\'alez Hern\'andez et al.\ 2005a). In this paper
we adhere to a distance of 1.2\,kpc.

No significant X-ray emission was detected before and just
after the 1969 flash of \object{Cen\,X-4} (Belian et al.\ 1972; see also Conner
et al.\ 1969). The 1-day average 3$\sigma$ detection limit of
{\it Vela 5B} is about 250 Uhuru Flux Units (see Priedhorsky \&
Holt 1987), which corresponds to about
6$\times$10$^{-9}$\,erg\,cm$^{-2}$\,s$^{-1}$
(2--10\,keV).\footnote{1 Uhuru Flux Unit (UFU) = 1 {\it Uhuru}
c\,s$^{-1}$ = 2.4$\times$10$^{-11}$\,erg\,cm$^{-2}$\,s$^{-1}$
(2--10\,keV), assuming a \object{Crab}-like spectrum (Forman et al.\
1978).} Using our light curves with 56\,hrs time resolution during the
$\simeq$40 days before the X-ray outburst we find
that the spread in values of the count rates is 32.4\,mCrab
(3--12\,keV). We, therefore, infer a 56\,hrs-average 3$\sigma$
upper limit of $\simeq$97\,mCrab (3--12\,keV) on the 
emission before the X-ray outburst. The error on the
average of the count rates over the 40 days is
$\simeq$6\,mCrab, from which we infer a 3$\sigma$ upper limit
on the emission over the whole 40-day time interval
of $\simeq$17\,mCrab (3--12\,keV). Assuming a \object{Crab}-like spectrum
(see, e.g., Kirsch et al.\ 2005) the above 3$\sigma$ limits translate to
2--10\,keV pre-flash fluxes of about
2$\times$10$^{-9}$\,erg\,cm$^{-2}$\,s$^{-1}$ and
3$\times$10$^{-10}$\,erg\,cm$^{-2}$\,s$^{-1}$, respectively. 
The non-detection of the source during 1977--1978 in
the HEAO-A2 sky survey corresponds to an upper limit of about
1$\times$10$^{-11}$\,erg\,cm$^{-2}$\,s$^{-1}$ (2--10\,keV; see
Kaluzienski et al.\ 1980), indicating that the source approached
quiescent values in between the two X-ray outbursts. The
quiescent flux observed long after the second X-ray
outburst is about 1--3$\times$10$^{-12}$\,erg\,cm$^{-2}$\,s$^{-1}$  
(0.5--10\,keV; see Rutledge et al.\ 2001).
It is, however, variable on long-term time scales ($\sim$40\%\ in 5
years; Rutledge et al.\ 2001), as well as on short-term time
scales (factor of about 3 in a few days and at a level of
$\sim$45\%\ rms down to about 100\,s; Campana et al.\ 1997, 2004). 

\section{Limits on pre-flash accretion rates}

\subsection{Limits from observations}

What is the level of accretion 
prior to the flashes, allowing them to occur? During none of the
flashes mentioned in the Introduction was
emission detected just before and after the flash,
so in principle no accurate measurement of the 
luminosity, and therefore accretion rate, can be made around
the time of these flashes. Nevertheless, the
observed upper limits may still give
us useful constraints on the accretion rate.

The ratio of (the upper limit to) the pre-flash emission and
the Eddington-limited flux provides an estimate of the
fraction of the Eddington limit at which a source is accreting
(see, e.g., Cornelisse et al.\ 2002a). For ease of comparison
between the observations and flash theory, we
therefore denote the observed X-ray luminosities in terms of
the Eddington luminosity, $L_{\rm Edd}$, and the accretion
rates, $\dot{M}$, in terms of the Eddington accretion rate, $\dot{M}_{\rm Edd}$. 
The Eddington accretion rate is defined as the accretion rate at 
which the corresponding accretion luminosity, $L_{\rm acc}$, equals 
$L_{\rm Edd}$.\footnote{$L_{\rm acc}$=$GM\dot{M}/R$$\simeq$1.3$\times$10$^{36}$\,$\dot{M}_{16}$($M/M_{\odot}$)(10\,km/$R$)\,erg\,s$^{-1}$,
where $L_{\rm acc}$ is the accretion
luminosity and where we used $\dot{M}$=10$^{16}$$\dot{M}_{16}$\,g\,s$^{-1}$
Note that this corresponds to an accretion efficiency of $\sim$0.2. 
For an accreting neutron star with a mass of 1.4\,M$_{\odot}$
and a radius of 10\,km, $L_{\rm acc}$$\simeq$1.8$\times$10$^{36}$\,$\dot{M}_{16}$\,erg\,s$^{-1}$.
See, e.g., Frank et al.\ (1992).}
The secondary star in \object{Cen\,X-4} is a late-type K3-K7 star in a
$\sim$15\,hr orbit with the neutron star (e.g., Casares et al.\ 2007, 
and references therein). We can fairly assume that the
secondary provides a solar mix of H and He; the metallicity is
only slightly over-solar (Gonz\'alez Hern\'andez et al.\ 2005b). 
For the other systems this is not clear.\footnote{The
secondary of \object{SAX\,J1808.4$-$3658} may be a hot
$\sim$0.05\,M$_{\odot}$ brown dwarf (Bildsten \&\ Chakrabarty 2001), 
while the secondary in \object{GRS\,1747$-$312}, lying in the
globular cluster \object{Terzan 6}, may be an ordinary subgiant 
(in 't Zand et al.\ 2003a), but no further details are known about
their composition, as well as for the other sources we consider in this paper.} 
Assuming that the metallicity of these sources does not deviate
much from solar,
$L_{\rm Edd}$$\simeq$2$\times$10$^{38}$\,erg\,s$^{-1}$ (see Sect.~2),
while correspondingly $\dot{M}$$_{\rm Edd}$$\simeq$10$^{18}$\,g\,s$^{-1}$ (see footnote 7,
using $L_{\rm acc}$=$L_{\rm Edd}$).

Since X-ray instruments provide information in a limited energy
range only, one has to correct the observed X-ray fluxes to
bolometric values. To determine the bolometric correction
factor one has to make assumptions on the spectrum 
outside the instrument's energy window. For the purpose of
this paper we employ a bolometric correction factor of about 2
(see in 't Zand et al.\ 2001, 2003b, 2007), whenever necessary.
We also assume that the pre-flash emission as well as the flash emission is isotropic.

For photospheric radius-expansion
flashes the peak bolometric black-body flux, $F_{\rm bb,peak}$, equals the 
Eddington flux, and, therefore, 
$L_{\rm x,bol}/L_{\rm Edd}$=$F_{\rm pers,bol}/F_{\rm bb,peak}$,
where $L_{\rm x,bol}$ and $F_{\rm pers,bol}$ refer to the (estimated) bolometric
pre-flash source luminosity and flux, respectively. 
When a flash does not show such an expansion phase, $F_{\rm bb,peak}$ represents a
lower limit to the Eddington flux, and, therefore, 
$L_{\rm x,bol}/L_{\rm Edd}$$<$$F_{\rm pers,bol}/F_{\rm bb,peak}$. The flux
values and corresponding luminosity information are shown in
Table~\ref{table_fluxes}. Note that the values for 
$L_{\rm x,bol}$/$L_{\rm Edd}$ are in the same range as the upper limits
derived on the emission of the burst-only sources
around the flashes (Cocchi et al.\ 2001, Cornelisse et al.\ 2002a, 
and references therein), where 
$L_{\rm x,bol}$/$L_{\rm Edd}$$\lesssim$0.002--0.02. 
The bolometric factor can be uncertain by about a factor of 2
(e.g., in 't Zand et al.\ 2007), and therefore, $F_{\rm pers,bol}$
and $L_{\rm x,bol}$ can be up to about a factor of 2 higher.
This does, however, not have an effect on our principal conclusions.

Assuming $L_{\rm x,bol}$/$L_{\rm Edd}$ equals
$\dot{M}$/$\dot{M}$$_{\rm Edd}$
(keeping in mind the uncertainties introduced by the
bolometric corrections, the assumption here of a radiation
efficiency of $\sim$0.2 and isotropic emission),
the above measurements suggest that $\dot{M}$ around the time of the flashes
in our sample is $\lesssim$10$^{-2}$\,$\dot{M}$$_{\rm Edd}$, i.e., 
at least about 2 orders of magnitude lower than what is typically observed when 
the systems are in outburst. 
We note here that the lowest measured accretion flux for an active burster 
is about 0.03\,$L_{\rm Edd}$ (in 't Zand et al.\ 2007).

\begin{table*}
\caption{Overview of the properties$^1$ of the observed flashes discussed in this paper.}
\begin{tabular}{cc@{}cl@{}c@{}c@{}c@{}c@{}c@{}c@{}c@{}}
\hline
\multicolumn{1}{c}{Source} &
\multicolumn{1}{c}{$F_{\rm bb,peak}$ (10$^{-8}$} &
\multicolumn{1}{c}{} &
\multicolumn{1}{c}{$F_{\rm pers}$} &
\multicolumn{1}{c}{} &
\multicolumn{1}{c}{$F_{\rm pers,bol}/F_{\rm bb,peak}$} &
\multicolumn{1}{c}{$L_{\rm x,bol}$ (10$^{36}$} &
\multicolumn{1}{c}{$\Delta t_{\rm last}$} &
\multicolumn{1}{c}{} &
\multicolumn{1}{c}{$\Delta t_{\rm next}$} &
\multicolumn{1}{c}{} \\
\multicolumn{1}{c}{} &
\multicolumn{1}{c}{erg\,cm$^{-2}$\,s$^{-1}$)} &
\multicolumn{1}{c}{} &
\multicolumn{1}{c}{(10$^{-10}$\,erg\,cm$^{-2}$\,s$^{-1}$)} &
\multicolumn{1}{c}{} &
\multicolumn{1}{c}{$\equiv$$L_{\rm x,bol}/L_{\rm Edd}$} &
\multicolumn{1}{c}{erg\,s$^{-1}$)} &
\multicolumn{1}{c}{(days)} &
\multicolumn{1}{c}{} &
\multicolumn{1}{c}{(days)} &
\multicolumn{1}{c}{} \\
\hline
\object{Cen\,X-4} & $\sim$120 & [1] & $<$20$^a$ (2--10\,keV) & [1] & $\lesssim$0.0033 & $\lesssim$0.7 & $>$40 & [1,2] & $\simeq$2 & [1,3] \\
\object{2S\,1711$-$339} & 3.0$\pm$1.0$^b$ & [4] & $<$0.7 (2--28\,keV) & [4] & $\lesssim$0.0047 & $\lesssim$1 & $\sim$340 & [4] & $\simeq$1440 & [5] \\
\object{GRS\,1747$-$312} & $\simeq$2.5$^c$ & [6] & $<$0.6 (0.1--200\,keV) & [6] & $\lesssim$0.0024 & $\lesssim$0.5 & $\simeq$26 & [6] & $\simeq$28 & [6] \\
\object{IGR\,J17473$-$2721} & $\simeq$11$^d$ & [7] & $<$1.7 (2--10\,keV) & [8] & $\lesssim$0.0031 & $\lesssim$0.4 & $\sim$960 & [9,10] & $\simeq$2 & [8,11] \\
\object{2S\,1803$-$245} & 3.1$\pm$0.7 & [12] & $<$2$^e$ (2--28\,keV) & [4] & $\lesssim$0.013 & $\lesssim$3 &  $\sim$7990 & [13] & $\simeq$7 & [12] \\
\object{SAX\,J1808.4$-$3658} & 25$\pm$2$^c$ & [14] & $<$3 (2--28\,keV) & [14] & $\lesssim$0.0024 & $\lesssim$0.5 & $\simeq$17 & [14] & $\simeq$543 & [15] \\
\hline
\multicolumn{11}{l}{\footnotesize $^a$\,{\it Vela 5B} 56\,hrs upper limit.}\\
\multicolumn{11}{l}{\footnotesize $^b$\,Highest peak flux among the flash sample in [4].}\\
\multicolumn{11}{l}{\footnotesize $^c$\,Photospheric radius-expansion flash.}\\
\multicolumn{11}{l}{\footnotesize $^d$\,Peak flux observed during another flash reported by [7].}\\
\multicolumn{11}{l}{\footnotesize $^e$\,No limits are quoted in [12]; we, therefore, assume the highest {\it BeppoSAX}/WFC flux upper limit from [4].}\\
\end{tabular}
\note{In the first two columns we give the observed flash bolometric
peak fluxes ($F_{\rm bb,peak}$) and the 3$\sigma$ upper limits on the
flux around the flashes ($F_{\rm pers}$). From
these we infer the flash peak luminosities in terms of the
Eddington luminosity ($L_{\rm Edd}$), assuming 
$L_{\rm x,bol}/L_{\rm Edd}$$\equiv$$F_{\rm pers,bol}/F_{\rm bb,peak}$ 
(see Sect.~3), where $F_{\rm pers,bol}$ and $L_{\rm x,bol}$ are the
bolometric pre-flash source flux and luminosity, respectively.
Assuming $L_{\rm Edd}$=2$\times$10$^{38}$\,erg\,s (see Sect.~2) we
derive an upper limit on $L_{\rm x,bol}$. In the last two columns
we provide the time since the last and to the next detected
X-ray outburst with respect to the time of the 
flash, $\Delta t_{\rm last}$ and $\Delta t_{\rm next}$,
respectively. References are given in brackets: 
[1] this paper, [2] Conner et al.\ (1969), 
[3] Belian et al.\ (1972), [4] Cornelisse et al.\ (2002a), 
[5] Markwardt \&\ Swank (2004), [6] in 't Zand et al.\ (2003b), 
[7] Altamirano et al.\ (2008), [8] Markwardt et al.\ (2008), 
[9] Grebenev et al.\ (2005), [10] Markwardt \&\ Swank (2005), 
[11] Del Monte et al.\ (2008), [12] Cornelisse et al.\ (2007), 
[13] Jernigan (1976), [14] in 't Zand et al.\ (2001),
[15] Marshall (1998).}
\label{table_fluxes}
\end{table*}

\subsection{Limits from theory}

The characteristics of a flash (such as duration
and peak luminosity) depend primarily on 
$\dot{M}$ and the composition of the accreted material, for a given neutron star mass
(e.g., Fujimoto et al.\ 1981, Fushiki \&\ Lamb 1987; see also
Peng et al.\ 2007, Cooper \&\ Narayan 2007, and references therein). 
In general, He ignites completely in a fraction of a second, while unstable hydrogen burning is
prolonged to about 100\,s through slow beta decays in the rp process (e.g.,
Cumming 2003, Woosley et al.\ 2004, Heger et al.\ 2007, Fisker et al.\ 2008). The duration
of a flash is a convolution of the burning process time and the cooling time. For 
$\dot{M}$$\lesssim$10$^{-2}$\,$\dot{M}$$_{\rm Edd}$
flashes occur deeper in the neutron star and the duration is mainly determined by the cooling time. 
At face value, the duration of the 1969 flash from \object{Cen\,X-4} is
consistent with burning of a H-rich layer with e-folding decay times close
to values found for the prototypical burster \object{GS\,1826$-$24} (e.g., Heger et al.\ 2007, 
Galloway et al.\ 2008). However, since the accretion rate at the time of \object{Cen\,X-4}'s flash
was low, the decay rate is likely to be limited also by the relatively large depth of the ignition.
Although the available literature on the theoretical
description of the flash behaviour at the low accretion
rates relevant to our sample (typically $\lesssim$10$^{-2}$\,$\dot{M}$$_{\rm Edd}$) 
is sparse and has only recently been 
developed in more detail (Peng et al.\ 2007, Cooper \&\ Narayan 2007), 
we can use it to infer additional constraints on the accretion rate around the
time of the observed flashes.

Below $\dot{M}$ of a few times 10$^{-3}$\,$\dot{M}$$_{\rm Edd}$, 
one expects mixed H/He flashes triggered by thermally
unstable H burning. The decay times of such flashes are on the
order of tens of seconds, due to the long waiting times
involved in $\beta$-decays. During the peak of these kinds of
flashes the fluxes are sub-Eddington, but 
are expected to exceed about 10\%\ of the Eddington limit.
Between a few times 10$^{-3}$\,$\dot{M}$$_{\rm Edd}$ and
10$^{-2}$\,$\dot{M}$$_{\rm Edd}$, unstable H burning does not
trigger unstable He burning and a layer of He is being built
up, which may eventually lead to an energetic (approaching or
even reaching the Eddington limit) and long flash.
These $\dot{M}$ limits depend, however, on the emergent flux
from the neutron star crust and whether sedimentation is
important or not. For example, a factor of 10 higher flux from
the crust results in $\dot{M}$ limits which are about a factor
of 10 lower, while including sedimentation has the effect of
increasing the $\dot{M}$ limits by about a factor of 2 (see
Peng et al.\ 2007, Cooper \&\ Narayan 2007). 

The minimum accretion rate below one expects no flashes
at all (i.e., completely stable burning all the time), 
is even less well-studied nor well-determined, but it is probably of
the order of 10$^{-5}$\,$\dot{M}$$_{\rm Edd}$ (see Fushiki \& Lamb 1987). 

It is important to realize that the above quoted $\dot{M}$
values from flash theory are presumed to sustain
for at least a few months (i.e., the thermal time scale of the
crust). If there were a temporary increase in $\dot{M}$ on a
shorter time scale, the temperature of the layer where the flash
originates may still be higher than would be expected from the lower 
long-term averaged $\dot{M}$, thus influencing the flash behaviour. 

\subsection{Comparing observations with theory}
\label{sect:comp}

We reiterate that, when comparing the observed limits on $\dot{M}$ with that inferred from flash theory,
one has to keep in mind the uncertainties and assumptions mentioned at the end of 
Sect.~3.1, as well as the uncertainties in the flash theory.
The minimum $\dot{M}$ below one expects no flashes,
translates to an accretion luminosity of a few times 10$^{33}$\,erg\,s$^{-1}$.
This is interestingly close, but above, to that observed for our sample
of sources, as well as other neutron star X-ray transients, in
quiescence (10$^{31-33}$\,erg\,s$^{-1}$; e.g., Cornelisse et
al.\ 2002a,b, 2007, Campana et al.\ 2004, Jonker et al.\ 2007, Heinke et al.\ 2007,
and references therein). 
The neutron stars in our source sample, therefore, must have been accreting at levels 
above that of quiescence, i.e., 
$\dot{M}$$\gtrsim$10$^{-5}$\,$\dot{M}$$_{\rm Edd}$. 
In this respect we note, however, the 
subluminous ($L_{\rm bb,peak}$$\sim$4$\times$10$^{36}$\,erg\,s$^{-1}$) 
flash seen from a source in M28 with a pre-flash luminosity of
$L_{\rm x,bol}$$\lesssim$10$^{33}$\,erg\,s$^{-1}$ (Gotthelf \&\ Kulkarni 1997), 
which may counter the suggestion.
For this source $L_{\rm x,bol}$/$L_{\rm Edd}$$\lesssim$5$\times$10$^{-6}$
(assuming solar type compositions), which is about a factor of 1000 lower than the values found
for the sources in our sample (Table~\ref{table_fluxes}).

The 1969 flash from \object{Cen\,X-4} was bright,
but we can not infer whether it had a photospheric
radius-expansion phase (i.e., reached the Eddington limit). It
could thus have been either a flash fueled by a
deep layer of nearly pure He or a long mixed H/He flash
triggered by an unstable H flash. The expected $\dot{M}$ is
thus $\lesssim$10$^{-2}$\,$\dot{M}$$_{\rm Edd}$, which is
consistent with the observed upper limit based on the Vela~5B
non-detection prior to the flash (see also 
Hanawa \&\ Fujimoto 1986). This also holds for the flashes seen
from \object{2S\,1711$-$339}, \object{IGR\,J17473$-$2721} and \object{2S\,1803$-$245}, as well as
those seen from the burst-only sources. The flashes
from \object{SAX\,J1808.4$-$3658} and \object{GRS\,1747$-$312} did reach the
Eddington limit and lasted on the order of minutes
(in 't Zand et al.\ 2001, 2003b). For these
sources we thus infer that the expected $\dot{M}$ was between
about a few times 10$^{-3}$\,$\dot{M}$$_{\rm Edd}$ and
10$^{-2}$\,$\dot{M}$$_{\rm Edd}$, roughly consistent with the observed
upper limits on the pre-flash source luminosity.
The expected $\dot{M}$ are 3--4 orders of magnitude
higher than that inferred from observations in
quiescence. 

One can make simple analytic estimates
how long such an enhanced accretion period has
to last in order to show a flash (assuming the flash is
produced from only the freshly accreted material, 
and that all of this fresh material is used during the flash). Burning H
to He to C (using a H mass fraction X=0.7 and He mass fraction
Y=0.3) gives roughly 5$\times$10$^{18}$\,erg\,g$^{-1}$, whereas
burning He to C (using X=0, Y=1) gives roughly
6$\times$10$^{17}$\,erg\,g$^{-1}$ (see, e.g., Bildsten 1998).
In the case of \object{Cen\,X-4}'s long 1969 flash, if it is due
to unstable mixed H/He burning, we need at least 10$^{21}$\,g
to get a flash fluence of more than 5$\times$10$^{39}$\,erg. If
$\dot{M}$ onto the neutron star is about
10$^{-4}$--10$^{-3}$\,$\dot{M}$$_{\rm Edd}$, one needs to
wait at least 115--12~days for a flash to 
occur. If the flash is due to unstable pure He
burning, one needs even more matter and one has to wait longer
(assuming the steadily burning H does not leave He behind).
Then the He mass needed is about 8$\times$10$^{21}$\,g, and the
neutron star needs to accrete at least an order of
magnitude longer than estimated above.
This enhanced accretion time scale grows inverse proportionally with mass accretion rate
for even lower values of $\dot{M}$.
Note that for $\dot{M}$$\lesssim$10$^{-4}$\,$\dot{M}$$_{\rm Edd}$, in the
mixed burning case, the time scale starts to become as long or even larger
than the thermal time scale of the crust.

Inspection of Fig.~1 (right panel) shows, in the case of
\object{Cen\,X-4}, no evidence of any increased X-ray activity (above
about 10$^{36}$\,erg\,s$^{-1}$), i.e., no enhanced accretion
onto the neutron star, above the {\it Vela 5B} detection level
during the 40~days prior to the 1969 outburst (see
also Conner et al.\ 1969). The other sources show comparable
limits on the existence of temporary increase in $\dot{M}$,
mainly based on observations with X-ray monitoring instruments 
such as the {\it RXTE}/ASM and {\it BeppoSAX}/WFC.
However, these limits (more than 3
orders of magnitude higher than the quiescent luminosity) are
not very constraining, and thus enhanced $\dot{M}$
could have occurred with maximum accretion luminosities of
about 5$\times$10$^{35}$--10$^{36}$\,erg\,s$^{-1}$, around the 
time of their flashes.

\subsection{Accretion during quiescence: conclusions}

We conclude that the presence of the flashes discussed here
require an enhancement of the accretion rate above that of pure quiescence, 
either temporary or on longer time scales.
In the first case the enhancement can occur in the phases just
before or after the bright outburst, or be the result of
a weak (presumably unobserved) outburst.

Indeed, Cornelisse et al.\ (2002a) already suggested that the
burst-only sources either accrete persistently at
``medium" $\dot{M}$ (with $L_{\rm x,bol}$$\simeq$10$^{34-35}$\,erg\,s$^{-1}$), 
or that their flashes occur during or after
faint outbursts (with $L_{\rm x,bol}$$\lesssim$10$^{36}$\,erg\,s$^{-1}$). 
Interestingly, one of
the burst-only sources, \object{XMMU\,J174716.1$-$281048}, was in
quiescence for some time and then started to accrete at low
levels with $L_{\rm x,bol}$$\simeq$6--10$\times$10$^{34}$\,erg\,s$^{-1}$ (see Del Santo
et al.\ 2007). Similarly, over the years \object{AX\,J1754.2$-$2754} has
been seen to accrete at low levels in between quiescence and
outburst (Jonker \&\ Keek 2008, and references therein).

Contrary to the burst-only sources, the systems 
\object{Cen\,X-4}, \object{IGR\,J17473$-$2721}, \object{2S\,1803$-$245},
\object{SAX\,J1808.4$-$3658}, \object{GRS\,1747$-$312} and \object{2S\,1711$-$339} 
have shown X-ray outbursts with 
$L_{\rm x,bol}$$>$10$^{36}$\,erg\,s$^{-1}$. Ten to eleven months after the
\object{SAX\,J1808.4$-$3658}'s 1998 outburst the system was in
quiescence (e.g., Campana et al.\ 2002), but it had shown
reflares after its 2000 and 2005 main outbursts
(Wijnands et al.\ 2001, Campana et al.\ 2008). 
Also \object{2S\,1711$-$339}
has been detected at low-intensity levels at various times
before and after its 1998/1999 X-ray outburst, with
fluxes of about a factor of 25 below the peak outburst
flux and about a factor of 3000 above quiescence (Cornelisse et
al.\ 2002a, Wilson et al.\ 2003, Torres et al.\ 2004a,b). 
Another neutron star X-ray transient LMXB, \object{4U\,1608$-$522}, also shows
low-intensity states (down by a factor of $\sim$10) after its
main outbursts, before going to quiescence (Keek et al.\ 2008). 
During its quiescence phase \object{SAX\,J1808.4$-$3658} was several times 
measured to be about 10 times brighter than at other times
(Campana et al.\ 2008). 
Quasi-persistent neutron star transients, such as \object{KS\,1731$-$260}
and \object{MXB\,1659$-$29}, have shown exponential decaying 
low-level intensity light curves after their main outburst 
over a period of several years
(from about 2--5$\times$10$^{33}$\,erg\,s$^{-1}$ to an apparent base level of 
2--5$\times$10$^{32}$\,erg\,s$^{-1}$; e.g., Cacket et al.\ 2006).\footnote{The smoothness 
of the decay of \object{KS\,1731$-$260} and \object{MXB\,1659$-$29} has been used 
by Cacket et al.\ (2006) as a counter argument for variations in $\dot{M}$ as the cause of
it; they favour a cooling neutron star. We note, however, that the decay light curves are not too
well sampled (effectively 5 observations over about 4 years).}
All of the above suggests the possibility of ongoing accretion
episodes after the main outburst, which may be universal
among the (neutron-star) X-ray transients. 

Although quiescence is reached some
time after the X-ray outbursts 
of \object{SAX\,J1808.4$-$3658}, \object{GRS\,1747$-$312} and \object{2S\,1711$-$339}, we
suggest that at the time of their flashes 
discussed in this paper $\dot{M}$
had not reached quiescent values yet.
For \object{2S\,1711$-$339} this was almost a year after the 
X-ray outburst (Table 1).
On the other hand, it could be that at the end of the quiescent
phase of \object{Cen\,X-4}, \object{IGR\,J17473$-$2721} and \object{2S\,1803$-$245},
$\dot{M}$ was enhanced as predicted by the DIM (see Sect.~4).
No evidence is found of a long-term increase of the 
accretion rate; the increase is of temporary nature. 
It is difficult to estimate the duration of this temporary enhancement, 
however, due to insufficient time coverage and sensitivity.

\section{Do flashes trigger X-ray outbursts, or is it the opposite?}

Flashes have an impact on the region around the
neutron star, i.e., the corona and/or the inner parts of the
accretion disk. This can be seen in those 
flashes where the residual X-ray emission during the
photospheric radius-expansion phase is {\it lower} than the
pre-flash emission (e.g., Molkov et al.\ 2000, Strohmayer
\&\ Brown 2002), and/or in those flashes where the
decay shows pronounced irregularities from canonical
exponential-like decay (e.g., van Paradijs et al.\ 1990,
Strohmayer \&\ Brown 2002, Molkov et al.\ 2005, in 't Zand et
al.\ 2005b, 2007; excluding those sources which are viewed at
high inclination). We here pose the question whether a flash 
can be influential enough to enhance the accretion onto the neutron star; 
in other words, was the 1969 \object{Cen\,X-4} flash the trigger of
\object{Cen\,X-4}'s subsequent X-ray
outburst (and similarly for \object{IGR\,J17473$-$2721} and
\object{2S\,1803$-$245})?

The answer depends on what causes outbursts in LMXBs.
It is generally accepted that the physical mechanism driving
these outbursts is analogous to that of dwarf novae 
(see Lasota 2001 for a review and references). The
main difference is that in LMXBs irradiation by the central
X-ray source strongly influences the disk's stability (van
Paradijs 1996, Dubus et al.\ 1999) and the properties of the
outburst cycle (Dubus et al.\ 2001).

According to the DIM, outbursts are driven by a thermal-viscous
instability which triggers heating fronts that bring an
initially `cold' disk into a `hot', high accretion-rate, state.
The hot outburst phase is ended by an inward-propagating cooling front
which brings the disk back to a cold state. During the
subsequent quiescent phase the disk fills up with matter until
it reaches the instability limit, which corresponds to the
critical accretion rate given by:
\begin{equation}
\dot{M}_{\rm crit}^{-}= 2.64\times10^{15}~\alpha_{0.1}^{0.01}~R_{10}^{ 2.58}~M_1^{-0.85}\,\rm g\,s^{-1}
\label{eq:mdotcrit}
\end{equation}
where $\alpha$=0.1$\alpha_{0.1}$ is the disk viscosity
parameter, $R$=10$^{10}{\rm cm}\, R_{10}$ the radius and
$M$=1\,$\rm M_{\odot}M_1$ the mass of the accreting object.
The superscript `$-$' refers to the maximum accretion rate for stable cold disks
(see Lasota et al.\ 2008).

During quiescence the accretion rate in the disk must
everywhere be lower than the critical value given by
Eq.~(\ref{eq:mdotcrit}). Therefore, for disks extending down to
the neutron star surface (or to the innermost stable circular
orbit) the resulting rates are ridiculously low, as
mentioned in the Introduction. 
However, for disks truncated at, say, $\sim$$10^9$\,cm, the accretion rate at
the inner disk's edge is about 7$\times$$10^{12}$\,g\,s$^{-1}$, i.e., 
$\sim$6$\times$10$^{-6}$\,$\dot{M}_{\rm Edd}$
(in agreement with observations of \object{Cen\,X-4} in
quiescence\footnote{Given the 1.2\,kpc distance and a bolometric correction factor of 
2 the observed X-ray luminosity in quiescence (see Sect.~2) corresponds to
$L_{\rm x,bol}$$\sim$3--10$\times$10$^{32}$\,erg\,s$^{-1}$. This 
leads to inferred accretion rates in quiescence of about 
2--5$\times$10$^{12}$\,g\,s$^{-1}$, or 2--5$\times$10$^{-6}$\,$\dot{M}_{\rm Edd}$.}). 
When the accretion rate reaches somewhere the critical 
value given by Eq.~(\ref{eq:mdotcrit}), an
outburst starts. In LMXBs the outbursts always start
near the inner edge of the (truncated) disk (see Dubus et al.\
2001). In the model of Dubus et al.\ (2001)
corresponding roughly to the parameters of \object{Cen\,X-4}, outbursts
start at $\dot{M}$$\sim$2$\times$$10^{14}$\,g\,s$^{-1}$ (see
their figure 19), i.e., at $\dot{M}$$\sim$2$\times$10$^{-4}$\,$\dot{M}_{\rm Edd}$,
which, (consistently) implies an accumulation time 
for a critical pile of flash fuel of several months to more than a 
year, depending on the sort of flash produced (see Sect. \ref{sect:comp}).

In outburst (when accretion rates can reach Eddington
limited values) the X-ray irradiation stabilizes the disk (van Paradijs
1996, Dubus et al.\ 1999), but in quiescence the X-ray
accretion luminosity is too low to affect the stability
properties. However, external disk irradiation (e.g., by a
flash) can affect the disk's properties by significantly
lowering the value of the critical density and accretion rate.
In general, the irradiation temperature can be written as:
\begin{equation}
\sigma T^4_{\rm irr} = {\cal C} \frac{L_{\sc x}}{4 \pi R^2},
\label{eq:ill}
\end{equation}
where ${L_{\sc x}}$ is the X-ray luminosity and ${\cal C}$
represents the fraction of the X-ray luminosity that heats up
the disk and contains information on the irradiation geometry,
the X-ray albedo and the X-ray spectrum. Based on observations
of irradiated disks in low-mass X-ray binaries, Dubus et al.\
(1999, 2001) found that in outburst, when $L_{\sc x}$ is the
accretion luminosity, ${\cal C}$$\approx$0.005.

In general, however, the effect of varying X-ray luminosity
can be mimicked by varying ${\cal C}$. 
The upper continuous line in figure 15 of Dubus et al.\ (2001)
shows the column-density profile of a disk just before the
onset of the outburst: in its inner parts the density
is very close to the critical one represented by a dotted line
in that figure.
This is the standard situation when irradiation in
quiescence is negligible because of low accretion luminosity.
The second (lower) dotted line in that figure corresponds to the case of a
disk irradiated in quiescence (${\cal C}$ multiplied by 100).
In this case the outburst would have
started at lower surface-density, i.e., earlier than in the case
of a non-irradiated disk. Can this represent the effect of
irradiation by a flash such as observed in \object{Cen\,X-4}?

As discussed in Sect.~3, the X-ray luminosity before the flash
was not higher than ${L_{\sc x}}$$\simeq$10$^{36}$\,erg\,s$^{-1}$. Therefore, the effect of
the flash can be roughly described by an increase in ${\cal C}$
by a factor of at least 100 which corresponds to the situation
represented at figure 15 of Dubus et al.\ (2001).
The numerical values of the surface-density and truncation
radius correspond to a 7\,$M_{\odot}$ accretor as it is the
only available example in Dubus et al.\ (2001). The relative amplitude
of the effect, however, does not depend on the mass of the compact
object. So, in principle, a 
flash can accelerate the start of an outburst.

However, since the flash lasted only about 10\,min, one has to check that the disk had the
time to modify its thermal structure, i.e., that it had
enough time to react to irradiation. The characteristic
thermal time scale of the disk is
$t_{\rm th}$$\sim$$1/(\alpha\Omega_{\rm K})$, where $\Omega_{\rm K}$ is 
the Keplerian angular velocity. Assuming an inner disk
radius at $\sim$10$^9$\,cm and $\alpha$=0.01 one gets $t_{\rm th}$$\sim$4\,min,
which is just right, but a larger radius
would preclude the flash significantly affecting the disk
structure. ($\alpha$ could have a value up to, say, 0.03, but
this would not help much).

As mentioned above, in the relevant non-irradiated 
model of Dubus et al.\ (2001), outbursts
start at $\dot{M}$$\sim$2$\times$$10^{14}$\,g\,s$^{-1}$.
From Eq.~(\ref{eq:mdotcrit}) this corresponds
to a radius of $\sim$4$\times$10$^9$\,cm. Therefore, it is
rather unlikely that the flash triggered the
outburst. If anything, the opposite is true:
according to the model, in the last two years preceding the
outburst the accretion rate increases to a level,
which may have provided the right conditions for the
flash to occur (see above). One should, however, keep
in mind that the exact value of the truncation radius depends
on an `evaporation' mechanism model.

\section{Summary}
\label{discussions}

The detection of a flash at the end of the quiescent phases in
\object{Cen\,X-4}, \object{IGR\,J17473$-$2721} and \object{2S\,1803$-$245}, in between
recurrent X-ray outbursts of \object{SAX\,J1808.4$-$3658} and
\object{GRS\,1747$-$312}, and after an X-ray outburst of
\object{2S\,1711$-$339}, shows that accretion was (temporarily) ongoing onto the
neutron star.
Such accretion must be at a level several orders
of magnitudes higher than that
inferred for quiescence, and the rates required are consistent
with the truncated disk model which predicts an accretion-rate 
enhancement before the onset of the outburst.
Therefore, it seems unlikely that crustal heating 
is the main source of luminosity in all of quiescence. Thus,
care must be taken when considering the properties of a cooling
neutron star when an outburst is over.

The DIM-predicted enhancement in accretion rate near the end of
quiescence of about 10$^{13}$ to
10$^{14}$\,g\,s$^{-1}$ might have provided the right conditions
for the flashes to occur in just before the
outbursts of \object{Cen\,X-4}, \object{IGR\,J17473$-$2721} and
\object{2S\,1803$-$245}. The uncertainties inherent in the DIM do not
allow us to test in detail the hypothesis that the flashes
triggered or accelerated the start of, the 
outbursts (althought we regard it as unlikely), nor the 2--7-day delay between the 
flash and the X-ray outburst.

\begin{acknowledgements}
We thank Jorge Casares for discussions regarding the secondary star of
\object{Cen\,X-4}, Randy Cooper regarding $\dot{M}$ limits from flash theory,
Masaru Matsuoka regarding the flash seen from \object{Cen\,X-4}
by {\it Hakucho}, Paul Gorenstein and Rick Harnden for discussions about the
{\it Apollo 15} X-ray burst event, Arne \AA snes, Bob Lin, David Sibeck and Matt Taylor
for discussions about solar wind particles and their interaction with the earth's
magnetosphere, and Peter Jonker for commenting on an earlier draft of the paper.
EK and JZ acknowledge support from the Faculty of the European Space Astronomy Centre (ESAC).
JPL was supported by the Centre National d'Etudes Spatiales (CNES).
\end{acknowledgements}

\appendix

\section{A flash from \object{Cen\,X-4} in quiescence recorded by {\it Apollo 15}?}
\label{apollo}

\subsection{{\it Apollo 15} observations}

The X-ray Fluorescence Spectrometer (1--$>$3\,keV), carried in the
Scientific Instrument Module of the Command and Service Module on the {\it Apollo 15} mission,
was mainly used for orbital mapping of the lunar surface composition
(see, e.g., Adler et al.\ 1972a,b, 1975).
Galactic X-ray observations were made during the trans-Earth coast for several periods of about an hour
(e.g., Adler et al.\ 1972b).
Among the various other instruments were the
Gamma-ray Spectrometer (500\,keV--30\,MeV; e.g., Arnold et al.\ 1972)
and the Alpha Particle Spectrometer (4.5--9.0\,MeV; e.g., Gorenstein \&\ Bjorkholm 1972).
The X-ray Fluorescence Spectrometer was pointed towards various constellations
over a period of about 3 days (1971, Aug 5--7) with an attitude stability within a degree
for approximately 1\,hr of observation (Adler et al.\ 1972b).

A large increase in the X-ray count rate for about 10\,min was recorded on UT 1971, August 5, 00:34
(Gorenstein et al.\ 1974).
The burst event occurred when the spacecraft was pointed towards the
Centaurus constellation, for a period of about 65\,min (Adler et al.\ 1972b, Gorenstein et al.\ 1974).
According to the Apollo 15 Flight Journal\footnote{See {\tt http://history.nasa.gov/ap15fj/}.}, 
it was aimed at \object{Centaurus A}.
The X-ray Fluorescence Spectrometer had all the known X-ray sources in Centaurus at that time
in the full-width conal field of view of 50$\degr$\footnote{The 
nominal field of view is about 60$\degr$ full-width at half-maximum
(see, e.g., Adler et al.\ 1972a,b); in practice, however, it proved out to be more
complicated (Adler et al.\ 1975).} (Gorenstein et al.\ 1974);
\object{Cen\,X-4} was about 21.5$\degr$ from the center of the field of view.

At the peak of the burst event the intensity was several times higher than \object{Sco\,X-1}, and
Gorenstein et al.\ (1974) noted that the light curve and strength of the burst event
was similar to that reported for the flash from \object{Cen\,X-4} by Belian et al.\ (1972).
Based on the above characteristics Gorenstein et al.\ (1974) suggested their
event to be a similar flash from \object{Cen\,X-4}.

The burst event was observed in the 1--3\,keV band, at $>$3\,keV, and at $>$\,7\,keV.
Unfortunately, there were no soft particle detectors on-board to discriminate
between local and Galactic radiation, so a passage through a region of
particles in cis-lunar space could not be ruled out. However, the background
level before and after the event was rather stable (Gorenstein et al.\ 1974).
No additional events were seen by the Gamma-ray Spectrometer and the Alpha Particle
Spectrometer, suggesting the event was of X-ray origin, supported by the sparse energy
information of the X-ray Fluorescence Spectrometer. In Sect.~\ref{background}
we elaborate further on this issue.

\begin{figure}[top]
\centering
  \includegraphics[height=.32\textheight,angle=-90]{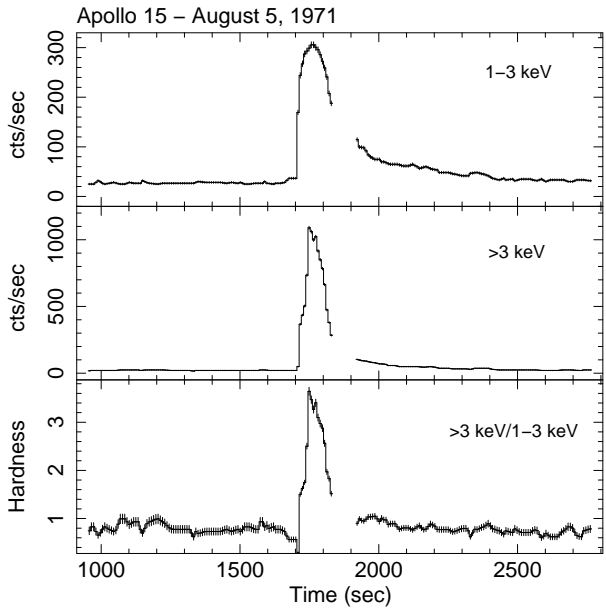}
  \caption{{\it Top and middle:} Light curves recorded by the X-ray Fluorescence Spectrometer
on-board {\it Apollo 15} in the 1--3\,keV band and at $>$3\,keV (taken from Gorenstein et al.\ 1974).
{\it Bottom:} Hardness curve derived from the ratio of the count rates at $>$3\,keV
and in the 1--3\,keV band. Data were recored every 8\,s.
The gap in the curves are due to observations of the on-board calibration source.
$t$=0\,s corresponds to UT 1971, Aug 5, 00:04.
}
\label{apollo15}
\end{figure}

We digitized the data shown in Figure 1 by Gorenstein et al.\ (1974), which
gives the events in the 1--3\,keV band and at $>$3\,keV (see Fig.~\ref{apollo15},
top panels). We assumed the uncertainties to be due to pure Poisson statistics, but note
that the systematic uncertainties are larger due to the pointing inaccuracy and
non-uniform collimator response.
We derived the hardness curve by taking the ratio of the
count rates at $>$3\,keV over the count rates in the 1--3\,keV band
(see Fig.~\ref{apollo15}, bottom panel).
Clearly, the event shows a fast-rise ($\simeq$50\,s from start to peak)
and exponential-like decay (decay time of $\simeq$100\,s and $\simeq$25\,s, respectively
in the 1--3\,keV band and at $>$3\,keV, from about $t$=1800\,s to 2000\,s).
During the rise to peak the source first hardens; subsequently, during the decay
the source softened. Indeed, this is typically seen for thermonuclear flashes
(see Sect.~2).

The event's peak rate is about 2450 and 8500\,cts per 8\,s
in the 1--3\,keV band and at $>$3\,keV, respectively, while \object{Sco\,X-1} varied between 1350--2050
and 2000--2750\,cts per 8\,s, respectively, as seen later on Aug 5 (see Adler et al.\ 1972b).
This correspond to a total peak rate about 2.3--3.3 times that of \object{Sco\,X-1}. The above quoted
rates are uncorrected for collimator response, therefore, if the event came from \object{Cen\,X-4}
the actual peak may be up to about a factor of 2 or so higher.
Measurements by the {\it RXTE} All-Sky Monitor (ASM) show that the daily average flux of \object{Sco\,X-1} is
roughly 11--13\,Crab (2--12\,keV). The Apollo 15 burst event peak flux is thus indeed
in the range observed for the flashes seen by {\it Vela 5B}
(Belian et al.\ 1972) and {\it Hakucho} (Matsuoka et al.\ 1980).
Although there were a couple of (currently) known flash sources (see, e.g., in 't Zand et al.\ 2004)
in the field of view of the X-ray Fluorescence Spectrometer at the time of the event,
they are all known to not show flashes like that seen from \object{Cen\,X-4}, 
and certainly not as bright.

\object{Cen\,X-4} was in quiescence around the time of the event (see Sect.~2).
During the event itself {\it Vela 5B} was operating, but did not look at \object{Cen\,X-4}'s region.
Prompted by the {\it Apollo 15} findings, we searched the {\it Vela 5B} data base for similar
flashes as seen in 1969 (see Sect.~2), but found none. Also the {\it RXTE}/ASM
data base on \object{Cen\,X-4} does not reveal any similar events above the level of about 0.5\,Crab.
It would be interesting to find out if similar events exist in other databases, such as
of {\it CGRO}/BATSE.

\subsection{Particle event or not?}
\label{background}

Energetic electron bursts at energies of keV to tens, or even hundredths, of
keV (but below about 500\,keV) are common around the Earth's magnetosphere.
They occur especially in the Earth's magnetotail in the anti-Sun direction, but also in
the magnetosheath which extends to $\pm$$\sim$30$\degr$ from the anti-Sun direction,
and around the bow shock (to $\pm$$\sim$45$\degr$ from anti-Sun);
see, e.g., Sarris et al.\ (1976), Baker \&\ Stone (1978), Meng et al.\ (1981).
They arise during geomagnetic activity.
If Apollo 15 was in these regions then there is a likelihood of such a burst;
either the spacecraft passed through a static particle region or such a region swept
across the spacecraft.

The position of the Sun and Moon on Aug 5, 1971, i.e., the longitude $\phi$ and latitude $\lambda$
in ecliptic coordinates were roughly $\phi$,$\lambda$=132$\degr$,0$\degr$ and 
$\phi$,$\lambda$=288$\degr$,$-$2$\degr$,
respectively. The Earth-Moon distance was around 374307\,km on that day.
According to the Apollo 15 Flight Journal$^{10}$ the spacecraft was about 15200\,km from the Moon
around the time of the Centaurus region observations, so we may assume it was
close to the position of the Moon. Apollo 15 was thus about 24$\degr$ from the
anti-Sun direction, placing it near the boundaries of the Earth's magnetotail.

The Interplanetary magnetic field was southward on Aug 5, 
1971\footnote{See {\tt http://omniweb.gsfc.nasa.gov}.},
favouring geomagnetic activity and the production of energetic particles in the Earth's
magnetosphere. This was particularly true for the first half day,
from UT 00:00--12:00. However, the activity was not exceptional
in any way, peaking from UT 03:00--04:00 and gradually decaying
the rest of the day. Further inspection\footnote{See 
{\tt http://swdcwww.kugi.kyoto-u.ac.jp/dstae/index.html}.}
also reveals a moderately strong geomagnetic substorm in progress at this time and
a weak geomagnetic storm.

If the burst event originated from increased geomagnetic activity, 
one would expect, however, to see more such events. Gorenstein et al.\ (1974) report
that, apart from the 1971, Aug 5 event, during a total about 20 hours of observing time of the 
X-ray Fluorescence Spectrometers onboard Apollo 15 and Apollo 16, two other events of two minute durations were 
seen, as well as one other event of longer duration. 
Also, as mentioned in Sect.~B.1, the X-ray flux was reported to be stable before and after the event.
This suggests that the Aug 5 burst event is a rather isolated event. 
Moreover, a changing slope of the spectrum of a particle burst in the magnetotail
is not likely to produce a dramatic change of spectral hardness, as observed in this
burst event.  We, therefore, conclude that the Aug 5 burst event has most probably a
celestial origin.

\end{document}